\newcommand{\Kp}{\mbox{$K^+$}}
\newcommand{\Km}{\mbox{$K^-$}}
\newcommand{\Kzero}{\mbox{$K^0$}}
\newcommand{\aKzero}{\mbox{$\bar{K^0}$}}
\newcommand{\Ks}{\mbox{$K^{0}_{s}$}}
\newcommand{\pp}{\mbox{$p$+$p$}}

\documentclass[12pt,a4paper,smallextended,epjc3]{svjour3}

\usepackage{amsmath}

\RequirePackage[T1]{fontenc}

\smartqed  

\RequirePackage{graphicx}
\RequirePackage{mathptmx}      
\RequirePackage{flushend}
\RequirePackage[numbers,sort&compress]{natbib}
\RequirePackage[colorlinks,citecolor=blue,urlcolor=blue,linkcolor=blue]{hyperref}

\AtBeginDocument{%
  \setlength{\oddsidemargin}{\dimexpr(\paperwidth-\textwidth)/2-1in}%
  \setlength{\evensidemargin}{\oddsidemargin}%
  \setlength{\topmargin}{%
    \dimexpr(\paperheight-\textheight)/2-\headheight-\headsep-1in}%
}

\begin{document}

\title{On the relation between $\Ks$ and charged kaon yields in proton-proton collisions}


\author{Joanna Stepaniak\thanksref{e1,addr1}
        \and
        Damian Pszczel\thanksref{e2,addr1} 
}

\thankstext{e1}{e-mail: joanna.stepaniak@ncbj.gov.pl}
\thankstext{e2}{e-mail: damian.pszczel@ncbj.gov.pl}

\institute{National Centre for Nuclear Research Otwock/Warsaw Poland\label{addr1}
}

\date{\today}

\maketitle

\begin{abstract}
A compilation of the experimental $\pp$ data of neutral and charged kaon production has revealed a discrepancy between the observed $\Ks$ yield and the average number of produced charged kaons $\Ks=(K^{+}+K^{-})/2$. This widespread relation holds only for a colliding system that is an uniform population w.r. to the isospin (i.e. that consists of an equal number of all members of a given isospin multiplet). This is not the case for $\pp$ collisions therefore one should not expect that it will describe the data accurately. A much better agreement is obtained with a model based on the quark structure of the participating hadrons.
\end{abstract}

\section{Introduction}\label{introduction}

The relation   between the yield of neutral kaons and the charged ones
is commonly expected to be in the form (see f.e.~\cite{Wroblewski:1985sz,Marek,NA49:2009wth}):\begin{equation} \label{eq1}
    \Ks = (\Kp + \Km)/2
    \end{equation}
\noindent 
where $\Ks$, $\Kp$, $\Km$ are the mean multiplicities of the corresponding kaon species.
This formula comes from the isospin symmetry.
However it is not valid for isospin asymmetric initial states.

In section~\ref{Rel}, we show a compilation of experimental data on neutral and charged kaon production in $\pp$ collisions for different $\sqrt{s}$ range.

In section~\ref{QPM}, we present a model based on the parton structure (valence and sea quarks) of different kaons and initial hadrons, the so-called Quark Parton Model (QPM).
Such model was used about 30 years ago in the design phase of the CNGS neutrino beam and kaons beams for NA31 experiment~\cite{Bonesini:2001iz}. 

In this paper, we show that this model works much better for $\pp$ interaction in a large energy range than Eq.~\ref{eq1}.

\section{Relation between charged and neutral kaon production}\label{Rel}

\begin{figure} 
    \begin{minipage}{\columnwidth}
    \centering
    \includegraphics[width=1.0\textwidth]{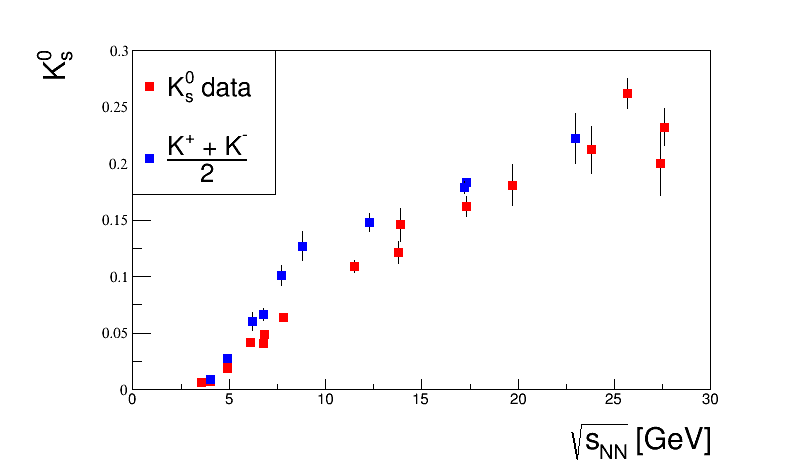}
    \end{minipage}
    \caption{Multiplicity of neutral kaons $\Ks$ and charged kaons $\frac{\Kp+\Km}{2}$ in inelastic $\pp$ interactions as a function of collision energy in the center of mass reference frame. Data points taken from~\cite{NA61SHINE:2021iay,Aduszkiewicz:2017sei,Marek}}
    \label{fig:k12k0}
    \end{figure}

The world data compilation of the average number of produced $\Ks$ per event in $\pp$ collisions is shown in Fig.~\ref{fig:k12k0} with red squares, whereas the blue ones correspond to the number of $\Ks$ extracted from $\Kp$ and $\Km$ production using Eq.~\ref{eq1}.

Blue squares are systematically located above the red squares. This shows that Eq.~\ref{eq1} is not adequate for $\pp$ collisions.

The relation between charged and neutral kaons can be derived
using the so called Smushkevich rule:
``For all particle involved in  isospin-conserving relation all members of isospin multiplets are produced in equal numbers \textbf{if and only if} the initial population is uniform''~\cite{Wohl:1982fu,Shmushkevich:1955,Dushin:1956,MacFarlane:1965wp}. It means that it is made up of equal numbers of the members of any multiplets involved. This is obviously not the case for $\pp$ collisions. 

The kaons form two isospin doublets: $\Kp-\Kzero$ and $\Km-\aKzero$.
Therefore e.g. in $\bar{p}$+$p$   collisions, which are obviously uniform we can expect the same number of $\Kp$ and $\Kzero$ as well as the same number of $\Km$ and $\aKzero$:
\begin{eqnarray}
\Kp = \Kzero \\
\Km = \aKzero  
\end{eqnarray}
therefore,
\begin{equation}
\Kp + \Km = \Kzero + \aKzero
\end{equation}
which leads to Eq.~\ref{eq1}.
This formula is not valid for the nonuniform isospin initial state $\pp$ with two members of $NN$ doublet with $I_{3}=+1/2$, therefore the observation of a discrepancy between Eq.~\ref{eq1} and $\pp$ data is not surprising.

In this case, when the initial population is nonuniform, a model based on the quark structure of the participating particles was proposed. 

\section{Derivation of Quark Parton Model formula}\label{QPM}

\begin{figure}[h] 
    \begin{minipage}{\columnwidth}
    \centering
    \includegraphics[width=1.0\textwidth]{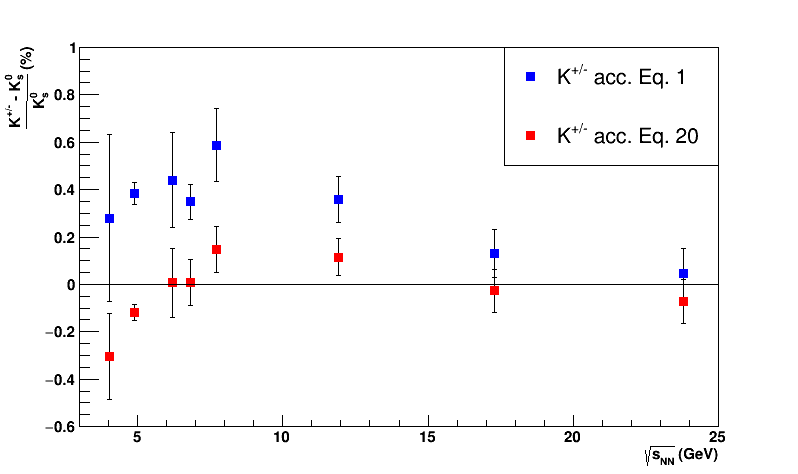}
    \end{minipage}
    \caption{Difference, normalized by $\Ks$, between a relation based on multiplicities of charged kaons given by the two models discussed in this paper (Eq~\ref{eq1} and Eq.~\ref{eqpp}) and $\Ks$ yields as a function of the center of mass energy of colliding protons.}
    \label{fig:tgraph}
    \end{figure}

\begin{figure}[h] 
    \begin{minipage}{\columnwidth}
    \centering
    \includegraphics[width=1.0\textwidth]{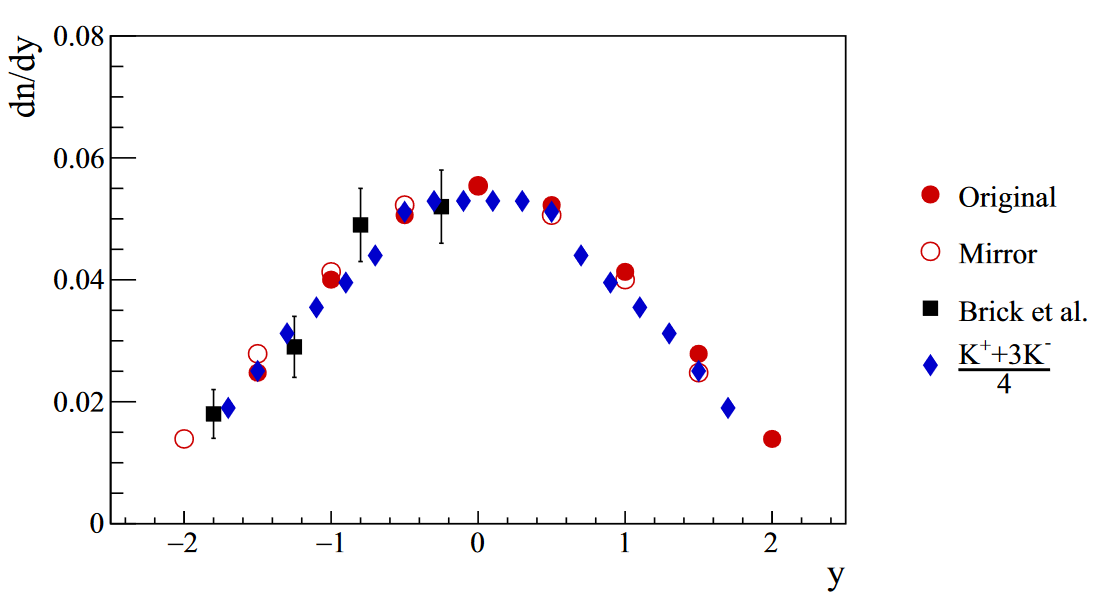}
    \end{minipage}
    \caption{Rapidity distribution $dn/dy$ of $\Ks$ mesons in inelastic $\pp$ interactions at $158$ GeV/c. Solid red-coloured circles correspond to the NA61/SHINE results (systematic uncertainties not shown on the plot), open circles are mirrored values, black squares represent results from Brick et al. at FNAL~\cite{Brick:1980vj} and blue full diamonds show results obtained from Eq.~\ref{eqpp} using charged kaon yields recently measured by NA61/SHINE at the same beam momentum~\cite{Aduszkiewicz:2017sei}. The figure is taken from~\cite{NA61SHINE:2021iay}.}
    \label{fig:dndy}
    \end{figure}

The quark structure of $K$ mesons is as follows:
\begin{align} 
\Kp & = u\bar{s}\\
\Km & = \bar{u}s\\
\Kzero & = d\bar{s}\\
\aKzero & = \bar{d}s\\
\Ks & = (\Kzero + \aKzero)/2 = ( d\bar{s} + \bar{d}s)/2
\end{align}
\noindent
In nucleon-nucleon interaction, $u$ and $d$ valence quarks exist in the initial state.
For example, $\Kp$ can be produced with either valence or sea quarks while $\Km$ with the sea quarks only:
\begin{align}
\Kp & = u_v\bar{s_s} + u_s\bar{s_s}\\
\Km & = \bar{u_s}s_s \\
\Kzero & = d_v\bar{s_s} + d_s\bar{s_s}\\
\aKzero & = \bar{d_s}s_s \\
\Ks & = (d_v\bar{s_s} + d_s\bar{s_s} + \bar{d_s}s_s)/2
\end{align}
\noindent
\break
Under the assumption that:

\begin{equation}\label{eq3}
u_s = d_s = \bar{u_s} = \bar{d_s} = \alpha
\end{equation}
\noindent
and
\begin{equation}\label{eq4}
s_s = \bar{s_s} = \gamma
\end{equation}
\noindent
we obtain:
\begin{align}
\Ks & = (d_v \gamma + \alpha\gamma + \alpha\gamma)/2\\
\Kp & = u_v \gamma + \alpha\gamma\\
\Km & = \alpha\gamma
\end{align}
\noindent
We still should multiply $d_v$ and $u_v$ by the number of corresponding valence quarks for the proton-proton, proton-neutron or neutron-neutron. After some simple arithmetic:
\begin{align}
p\textrm{+}p: \Ks &= \frac{1}{4}(\Kp + 3\Km)\label{eqpp}\\
n\textrm{+}n: \Ks &= \Kp\\
p\textrm{+}n: \Ks &= \frac{1}{2}(\Kp + \Km)
\end{align}
\noindent
We compare the number of $\Ks$ from the charged kaons average model:
$$\Ks = \frac{1}{2}(\Kp + \Km)$$
with the one calculated using QPM:
$$\Ks = \frac{1}{4}(\Kp + 3\Km)$$ as a function of $\sqrt{s}$. The results are shown in Fig.~\ref{fig:tgraph}.
\newline
\newline
\noindent
Only such data points were shown for which  the $\Kp$, $\Km$ and $\Ks$ multiplicities were measured at the same or close energy. One can see that the QPM formula leads to  much better agreement with the line drawn at zero.
Only the statistical uncertainty is provided for the second point at $4.9$ GeV.
\newline
\noindent    
Additionnaly, there is indication that QPM is describing $\pp$ data well in different rapidity bins as shown in Fig.~\ref{fig:dndy}.
\section{Summary}

We have noticed that the widespread relation between the average number of produced charged and neutral kaons holds only for a colliding system that is an uniform population w.r. to the isospin (i.e. that consists of an equal number of all members of a given isospin multiplet). This is not the case for $\pp$ collisions therefore one should not expect that Eq.~\ref{eq1} will describe the data accurately. 

We have shown that a better agreement with world data for a wide energy range is obtained if one uses a relation derived from  simple considerations about the quark structure of kaons and nucleons (Eq.~\ref{eqpp}).

\begin{acknowledgements}
The authors would like to thank Marek Gaździcki for his numerous comments and help during the preparation of this publication. We gratefully acknowledge insightful discussions with Andrzej Rybicki, Ludwik Turko, Wojciech Bryliński, Marjan Cirkovic and other members of the NA61/SHINE collaboration.
This work was supported by the National Science Centre Poland - the Norwegian Financial Mechanism (grant nr. 2019/34/H/ST2/00585) and  the Polish Minister of Education and Science (contract No. 2021/WK/10).
\end{acknowledgements}



\bibliographystyle{unsrtnat}
\bibliography{QM_paper}
    






\end{document}